\documentclass[10pt,conference]{IEEEtran}

\usepackage{algorithmicx}
\usepackage{tikz}
\usetikzlibrary{calc,angles,positioning,intersections,quotes,decorations.markings}
\usepackage{pgfplots}
\pgfplotsset{compat=1.16}
\tikzset{
	font={\fontsize{14pt}{12}\selectfont}}
\usepackage[noend]{algpseudocode}
\usepackage{amsmath,amsthm} 
\usepackage{array}
\usepackage{caption}
\usepackage{subcaption}
\usepackage{graphicx}
\usepackage{xcolor,colortbl}

\usepackage{epstopdf}
\usepackage{setspace}
\usepackage{ctable}
\usepackage{romannum}
\usepackage{enumerate}
\usepackage{ragged2e}
\usepackage{multirow}
\usepackage{tabularx,tabulary}
\usepackage{nicefrac}		
\usepackage[flushleft]{threeparttable}
\usepackage[ruled,vlined]{algorithm2e}
\usepackage{hyperref}
\SetKwRepeat{Do}{do}{while}
\usepackage{rotating}
\usepackage{tabularx,ragged2e,booktabs}
\usepackage[justification=centering]{caption}
\usepackage{listings}
\usepackage{amssymb}
\usepackage{pifont}
\usepackage[most]{tcolorbox}
\definecolor{backcolour}{rgb}{0.95,0.95,0.91}
\usepackage{lstlinebgrd}
\lstdefinestyle{customc}{
escapechar=?,
  abovecaptionskip=-5pt,
  breaklines=true,
  numbers=left, 
  floatplacement=t,
  frame = TRBL,
  xleftmargin=\parindent,
  language=C,
  showstringspaces=false,
  basicstyle=\footnotesize\ttfamily,
  keywordstyle=\bfseries\color{green!40!black},
  commentstyle=\itshape\color{purple!40!black},
  identifierstyle=\color{blue},
  stringstyle=\color{orange},
}
\newcolumntype{C}[1]{>{\Centering}m{#1}}

\usepackage{mathtools}
\usepackage{verbatim}

\usepackage{soul}
\usepackage{xcolor}

\usepackage{acronym}
\acrodef{FuCE}{fuzzing and concolic execution}
\acrodef{HLS}{high-level synthesis}
\acrodef{IP}{intellectual property}
\acrodef{DUT}{design-under-test}
\acrodef{COTS}{commercial-off-the-shelf}
\acrodef{SoC}{system-on-chip}
\acrodef{CGF}{coverage-guided greybox fuzzing}
\acrodef{AFL}{American fuzzy lop}
\acrodef{S2E}{symbolic executer}
\acrodef{HDL}{hardware description language}
\acrodef{SAT}{Satisfiability}
\acrodef{AFL-SHT}{AFL-SHT}
\acrodef{SCT-HTD}{SCT-HTD}
\acrodef{BCOV}{Branch coverage}
\acrodef{LCOV}{Line coverage}
\acrodef{FCOV}{function coverage}
\acrodef{RTL}{register-transfer level}
\acrodef{IR}{intermediate representation}
\acrodef{SHT}{Synthesizable Hardware Trojan}
\acrodef{CTSC}{Concolic Testing of SystemC Designs}
\acrodef{SESC}{Symbolic Execution of SystemC Designs}

\usepackage{amssymb}
\usepackage{pifont}
\newcommand{\solution}{GreyConE}

\hypersetup{
    colorlinks,
    linkcolor={red!50!black},
    citecolor={blue!50!black},
    urlcolor={red!50!black}
}
\usepackage{comment}

\ifCLASSINFOpdf
\else
\fi
\begin{document}
\bstctlcite{IEEEexample:BSTcontrol}
%
\title{\solution: Greybox fuzzing+Concolic execution guided test generation for high level design}


\author{
\IEEEauthorblockN{Mukta Debnath\IEEEauthorrefmark{1}\IEEEauthorrefmark{3}, Animesh Basak Chowdhury\IEEEauthorrefmark{2}\IEEEauthorrefmark{3}, Debasri Saha\IEEEauthorrefmark{4} and Susmita Sur-Kolay\IEEEauthorrefmark{1}
}
\IEEEauthorblockA{\IEEEauthorrefmark{1} Advanced Computing and Microelectronics Unit, Indian Statistical Institute, Kolkata, India\\
\IEEEauthorrefmark{2}Centre for Cybersecurity, New York University, United States\\
\IEEEauthorrefmark{4}A. K. Choudhury School of IT, U. of Calcutta, India\\
\IEEEauthorrefmark{3}Equal contribution
}
}

\markboth{Journal of \LaTeX\ Class Files,~Vol.~14, No.~8, August~2015}%
{Shell \MakeLowercase{\textit{et al.}}: Bare Demo of IEEEtran.cls for IEEE Transactions on Magnetics Journals}
%





\IEEEtitleabstractindextext{%
\begin{abstract}
Exhaustive testing of high-level designs pose an arduous challenge due to complex branching conditions, loop structures and inherent concurrency of hardware designs. Test engineers aim to generate quality test-cases satisfying various code coverage metrics to ensure minimal presence of bugs in a design. Prior works in testing SystemC designs are time inefficient which obstruct achieving the desired coverage in shorter time-span. We interleave greybox fuzzing and concolic execution in a systematic manner and generate quality test-cases accelerating test coverage metrics. Our results outperform state-of-the-art methods in terms of number of test cases and branch-coverage for some of the benchmarks, and runtime for most of them.

\end{abstract}

\begin{IEEEkeywords}
SystemC, High-level Synthesis, Greybox fuzzing, Symbolic Execution, Concolic Execution, Code Coverage
\end{IEEEkeywords}}

\maketitle

\IEEEdisplaynontitleabstractindextext

\IEEEpeerreviewmaketitle

\thispagestyle{plain}
\pagestyle{plain}


\section{Introduction}
\label{label:intro}


\IEEEPARstart{S}{ystemC} is the latest \textit{de-facto} standard for early-stage development of hardware designs having a complex design architecture. Detecting functional and security bugs at early stage design development is crucial to eliminate functional inconsistencies and vulnerabilities. A SystemC design provides cycle-accurate behavioural modeling of a given specification.  Therefore, testing SystemC design focuses more on functional level correctness, inconsistency in specifications and unknown (possibly erroneous) behaviour rather than defects arising from low-level hardware. Hardware engineers prefer to check logical inconsistencies in early-stage design as defects caught post manufacturing stage are costly. This motivates verification and test engineers to develop a good test generation framework for detecting bugs for designs at higher abstraction level.

Verification and testing SystemC designs (and, in general high level designs) for bug detection have a rich literature~\cite{FormVerPetriNet,ModCheckSysCAuto,symba,Rolf1}. Prior works adopt two mainstream approaches to tackle the problem: (a) formal techniques, e.g. model checking~\cite{ModCheckSysCAuto} and symbolic execution~\cite{symba}; and (b) simulation-based techniques such as black-box testing, white-box~\cite{DART,sage} and grey-box~\cite{afl,snpfuzzing} fuzzing. There are a plethora of works that have critically identified the issues like \textit{scalability} and \textit{exhaustive testing}, and addressed them by proposing a hybrid approach --- \textit{concolic testing}. A very recent work~\cite{2018_concolic_sysC} has applied concolic testing to alleviate the problem of scalibility arising from symbolic execution. Although the approach is quite promising, we ask two unaddressed questions:
\begin{enumerate}
    \item Concolic testing relies heavily on initial test-cases. How can one generate better quality test-cases (in a minimalistic time frame) such that concolic execution can focus only on ``hard-to-reach" state of design?
    \item Guided fuzzing-based techniques quickly explore the design space but often get stuck in complex branch conditions. How can one mitigate this problem in fuzzing?
\end{enumerate}

In this work, we propose \solution{}: an end-to-end test-generation framework for high-level designs outputting high-quality test-cases and quickly achieves better coverage metrics. In short, we make the following contributions:
\begin{enumerate}
    \item We developed \emph{\solution{}} by interleaving greybox fuzzing and concolic execution. We leverage the power of greybox fuzzing by quickly covering "easy-to-reach" states first and pass on ``hard-to-reach" state to concolic engine.
    \item \emph{\solution{}} outperforms state-of-the-art Greybox fuzzer (AFL~\cite{afl}) and concolic test generation (S2E~\cite{s2e}) in terms of branch coverage achieved and runtime for many of the benchmark SystemC designs. 
\end{enumerate}

The rest of the paper is organized as follows: Section~\ref{sec:background} outlines the background and  prior related works. In Section~\ref{sec:framework}, we present our proposed framework and show the efficacy of results in Section~\ref{sec:results}. Section~\ref{sec:conclusion} has the concluding remarks.

\section{Background}
\label{sec:background}

\subsection{Test-case generation for SystemC Designs}
\label{label:background_overview}
Recent works like SESC \cite{2016_concolic} and CTSC \cite{2018_concolic_sysC} use symbolic and concolic execution respectively to test SystemC designs. SESC  adapted symbolic execution with significant code coverage but limited to only a subset of SystemC features. Later, CTSC covered most SystemC semantics but failed to improve efficacy in terms of performance and scalability compared to SESC. Further, both the work depends on the availability of initial test-cases for the symbolic engine.



\subsection{Greybox fuzzing} Fuzz testing is a well known technique for detecting bugs in software. Greybox fuzzing~\cite{afl,chowdhury2019verifuzz} generates interesting test-cases using genetic algorithm on instrumented executable.  Instrumentation injects markers in the code at every basic-block which can track post compilation and execution whether a test-case has reached the marker location. A fitness function is used in order to evaluate the quality of a test-case. Typically, greybox fuzzing is used to improve branch-pair coverage of the design. A test-case is ``interesting", if it covers a previously unexplored basic block, or covers it for a unique number of times. The fuzzer maintains a history-table for every basic block covered so far, and retains interesting test cases for further mutation/crossover. The test generation process halts once the used-defined coverage goal is achieved. 
Popular \ac{CGF} engines like AFL~\cite{afl} are able to detect countless hidden vulnerabilities in well-tested software.

\subsection{Symbolic and Concolic Execution} 
\begin{figure}[t]

\resizebox{0.2\textwidth}{!}{
\begin{minipage}[c][0.8\width]{
  0.23\textwidth}
  \centering
\begin{lstlisting}[style=customc,label={label:symbeg}]
int foo(int i, int j){
//i and j symbolic variables
  int x = 0, y = 0, z;
  z = 2*j;
  if(z == i){
    if (j < 5){
         x = -2; y = 2;
     }
     else 
      { x = 2; y = 2;}
  }
  else 
   { x = 2; y = 2;}
  return 0;
}

\end{lstlisting}
\end{minipage}
}
\hfill
  {
\begin{minipage}[c][1\width]{
  0.23\textwidth}
  \centering
  \includegraphics[width=1\textwidth]{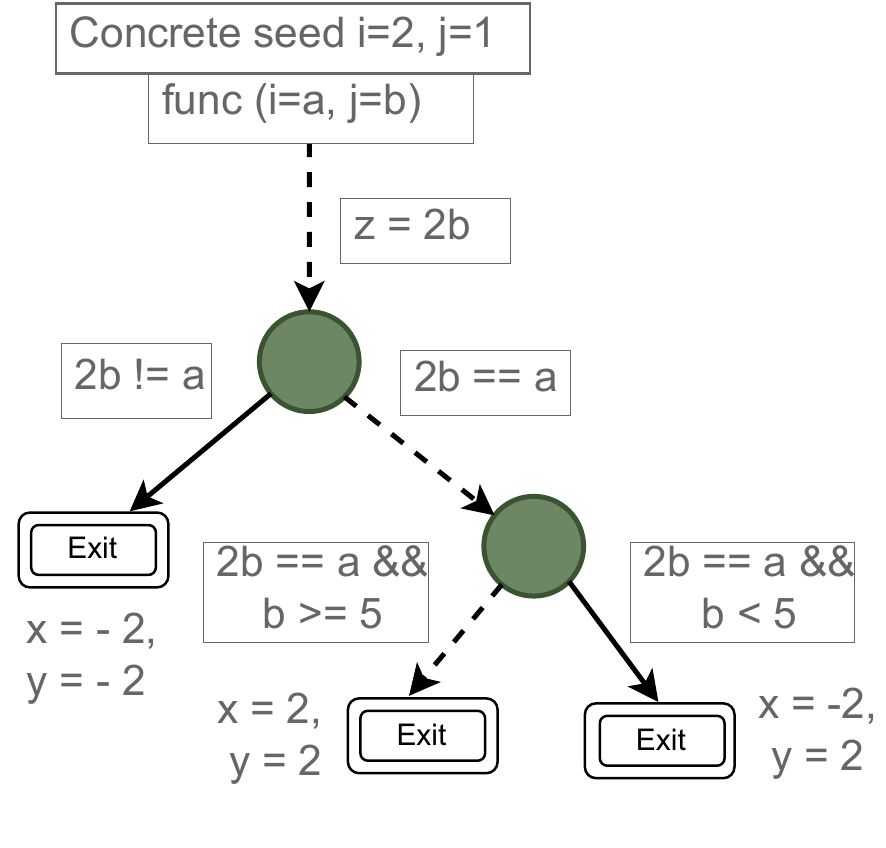}
\end{minipage}}
\caption{(a) Example code snippet. (b) Symbolic and concolic execution flow}
\label{fig:symb-conc}
\end{figure}

\textit{Symbolic} execution is a formal technique for generating quality test-cases. It typically assigns program inputs as symbolic values (as opposed to concrete values) and forks two threads at each condition; a boolean formula evaluating to True and the negated formula. Thus, symbolic execution forms an execution tree (\autoref{fig:symb-conc}(b)). At each branching condition, a SAT/SMT solver is invoked to generate test-case which in turn finds a set of concrete values assigned to symbolic variables reaching the branching condition. The size of the execution tree grows exponentially in terms of branching conditions resulting in state-space explosion.

One major drawback of symbolic execution is treating all program inputs as symbolic variables. This results in complex boolean formulae within a few nested conditions resulting in longer runtime for generating test-cases. \textit{Concolic} execution solves this problem by assigning concrete values to a set of program inputs for quicker search space exploration. This helps covering the branching conditions which are ``easy-to-reach", and pass on the simplified formulae for ``hard-to-reach" conditions to SAT/SMT solver. Post executing the program with concrete values, constraints at each condition are negated and solved to generate a new test case covering the unexplored path. The path constraints generated have reduced the number of clauses and variables aiding solvers to penetrate deeper program states quickly.

We demonstrate concolic execution with a simple example (\autoref{fig:symb-conc}a). A set of program inputs are treated as symbolic variables; the inputs $i$ and $j$ have symbolic values $i=a$ and $j=b$. We choose a random concrete input ($i=2, j=1$) and obtain the execution trace (dotted lines, \autoref{fig:symb-conc}b). The alternative path constraints to be explored symbolically are collected along the execution path guided by concrete inputs, forking the side branches. At each condition, constraints are negated and solved to generate new test cases. The concolic engine terminates after all conditions are covered. In the next section, we present \textit{Greycone}, which interleaves concolic execution and greybox fuzzing to accelerate test-case generation.



\begin{figure*}[h]
\begin{center}
\includegraphics[width=1.8\columnwidth]{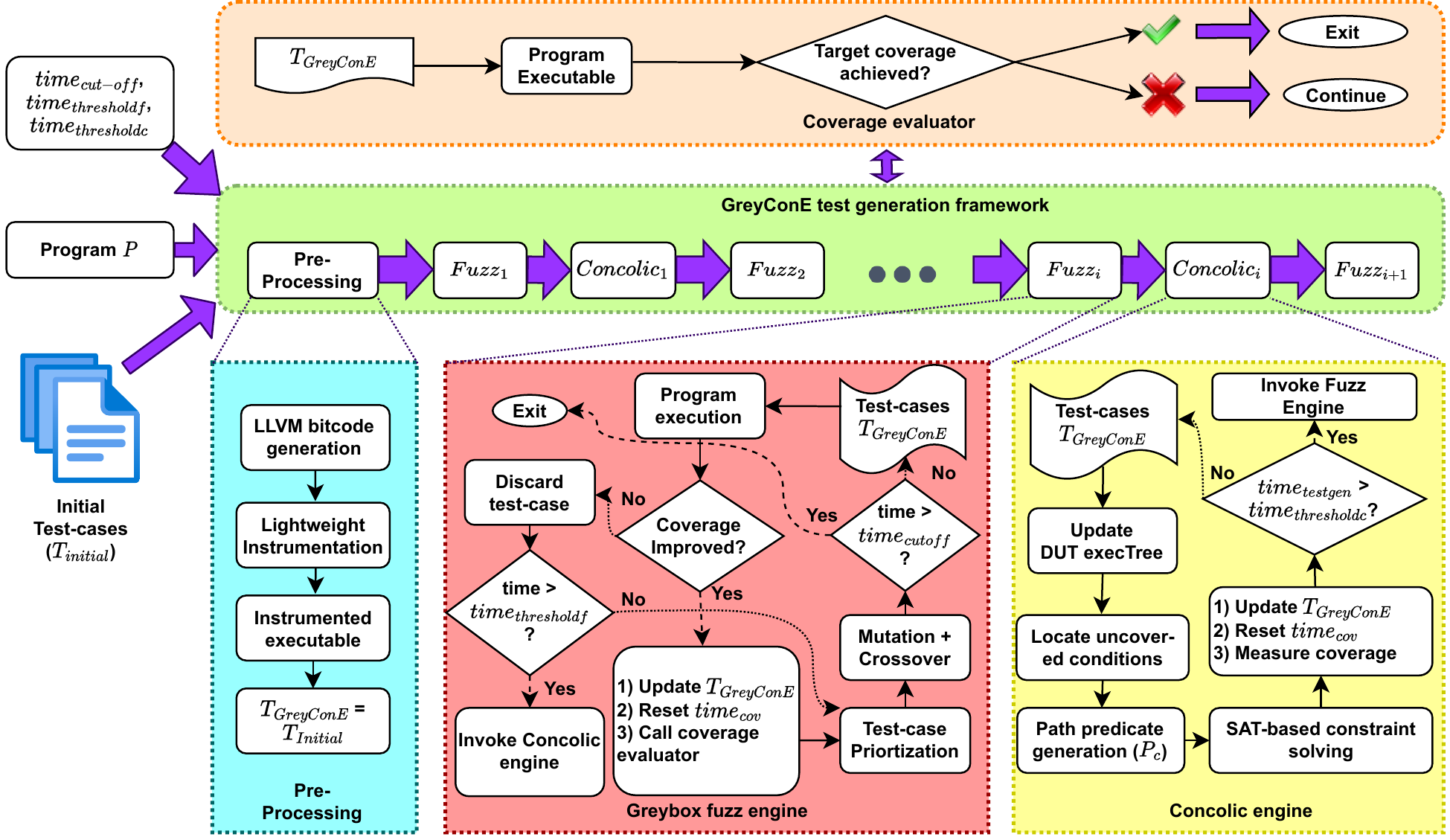}
\end{center}
\caption{\emph{GreyConE} test generation framework} 
\label{fig:flow_chrt}
\end{figure*}


{\small
\begin{algorithm}[!ht]
\footnotesize
\SetAlgoLined
\KwData{$DUT_{ins-exe}$, $T_{initial}$, $time_{cut-off}$} 


\KwResult{{$T_{fuzzed}$}}       
$T_{fuzzed} \gets T_{initial}$ \\
\While{time $ \leq time_{cut-off}$}{
    \For{$\tau \in T_{fuzzed}$} 
   {
        $K\gets$ CALCULATE\_ENERGY($\tau$)\\
        \For{$i \in \{1,2,\dots,K\}$}{
            $\tau'\gets$ MUTATE-SEED($\tau$) \\ 
            \If{IS-INTERESTING($DUT_{ins-exe}, \tau'$)}
            {
            $T_{fuzzed}\gets$ $T_{fuzzed}\cup \tau'$
            }
        }
        }
    }
    \Return $T_{fuzzed}$
\caption{FUZZER($DUT_{ins-exe}$, $T_{initial}$)}
\label{algo:AFL_FUZZ}
\end{algorithm}
}




\section{GreyConE framework}
\label{sec:framework}
The workflow of \emph{GreyConE} (Fig.~\ref{fig:flow_chrt})
comprises (1) Greybox fuzzing, (2) Concolic execution, and 3) Coverage evaluator.
\subsection{Generating instrumented binary} First, we convert a high-level hardware design to a low-level intermediate representation (IR) in the form of LLVM bitcode \cite{LLVM}. The compiler inserts a marker at the top of every basic block in the IR to generate an instrumented executable which is passed to both greybox fuzzing engine and concolic engine to generate test-cases and track uncovered basic blocks.

\subsection{Greybox fuzzing by AFL}
We outline the overall flow of Greybox fuzzing by \ac{AFL}~\cite{afl} in Algorithm \ref{algo:AFL_FUZZ}. The instrumented executable $DUT_{ins-exe}$ and a  test-set $T_{initial}$ are fed to the fuzzing framework. 
The function \textit{CALCULATE-ENERGY} assigns energy to every test-case of $T_{initial}$ based on runtime behaviour, e.g. execution time and obtained coverage. It assigns more energy to a test-case with faster execution time, covering more branches and penetrating deeper code segments. \ac{AFL} uses $T_{initial}$ to perform \textit{deterministic} and \textit{havoc} mutations to generate newer test-cases (\textit{MUTATE-SEED}). \ac{AFL} uses branch-pair coverage as a fitness metric to determine the quality of a test-input. For each branch-pair, it maintains a hash-table entry to record hit counts. It retains a test-input for further exploration if it covers an unseen new-branch pair or has unique hit-counts on already covered branch-pair (\textit{IS-INTERESTING}). The fuzz engine retains interesting test-cases for further exploration. The algorithm terminates post covering all branching conditions or reaching a user-defined $time_{cut-off}$. \ac{AFL} maintains all interesting test-inputs in the queue $T_{fuzzed}$.

\subsection{Concolic testing by S2E}
For concolic testing, we employ \ac{S2E}~\cite{s2e} which has two main components: 1) a virtual machine based on QEMU~\cite{qemu} and 2) a symbolic execution engine based on KLEE~\cite{klee}. The interesting inputs in the design can be marked as symbolic using S2E API \textit{s2e\_make\_smbolic()} function. 
It executes the $DUT_{ins-exe}$ with concrete test-inputs from $T_{initial}$ (\textit{CONC-EXEC}) generating concrete execution traces. It maintains the state of execution tree $DUT_{execTree}$ and identifies uncovered branch-pairs using $T_{initial}$.  \textit{COND-PREDICATE} constructs the path constraints for the uncovered branch pair, forks new thread and invokes SAT-solver (\textit{CONSTRAINT-SOLVER}) to generate the test-cases. It terminates either after a user-defined timeout is reached or covers all branch-pairs of $DUT_{execTree}$. $T_{concolic}$ stores the test cases generated by the concolic engine. We outline a typical concolic execution approach in Algorithm \ref{algo:concolic_execution}. 

{\small
\begin{algorithm}[t]
\footnotesize
\SetAlgoLined
\KwData{$DUT_{ins-exe}$, $T_{initial}$}
\KwResult{$T_{concolic}$}
$DUT_{execTree} \gets \phi$ \\
\For{$\tau \in T_{initial}$}
{
    $P_{trace} \gets $ CONC-EXEC($DUT_{ins-exe}$,$\tau$)\\
    $DUT_{execTree} \gets DUT_{execTree} \cup P_{trace}$
}
\For{uncovered cond $c \in DUT_{execTree}$}
{
   $p_{c} \gets$ COND-PREDICATE($c$) \\
   $t_i$ $\gets$ CONSTRAINT-SOLVER($p_c$)\\
    $T_{concolic}$ $\gets$ $T_{concolic}$ $\cup$ $t_i$
    
} 
\Return $T_{concolic}$
\caption{CONCOL-EXEC($DUT_{ins-exe}$,$T_{initial}$)}
\label{algo:concolic_execution}
\end{algorithm}
}

\subsection{Interleaved fuzzing and concolic testing} We interleave greybox fuzzing and concolic execution extenuating the problems associated with standalone techniques. Concolic engine usually performs depth-first search for search-space exploration. A set of random test-cases lead to frequently invoking the SAT/SMT solver for generating test-cases to cover unexplored conditions in $DUT_{execTree}$. We perform fuzzing of the DUT to avoid this huge slowdown and invoke concolic engine for ``hard-to-cover" scenarios with the fuzzer generated test-cases.
As shown in Fig.~\ref{fig:flow_chrt}, we first perform lightweight instrumentation on all conditions of \ac{DUT} and generate an instrumented executable. We start our fuzz-engine (\textit{FUZZER}) with a set of initial test-cases. The fuzz-engine  generates interesting test-cases using genetic algorithm and explores various paths in the design. Once 
a user-defined time period $time_{threshold_f}$ expires, we invoke the concolic engine (\textit{CONCOL-EXEC}) with fuzzed test-cases for unseen path exploration.
\textit{CONCOL-EXEC} identifies uncovered conditions in $DUT_{execTree}$ and forks new threads for symbolic execution on such conditions using depth-first search. The concolic engine generates new test-cases satisfying complex conditional statements.
We limit the runtime of concolic execution engine to avoid scalability bottleneck by $time_{threshold_c}$, that monitors the time elapsed since the last test case generated. Test-cases generated by concolic engine are fed back to the fuzz engine for quicker exploration once the ``hard-to-cover" conditions are explored. 
This process halts when either a user-defined target coverage is achieved or a user-defined time limit $time_{cutoff}$ is reached.

\section{Experimental setup and design}
\label{sec:results}

\subsection{Experimental setup}

We implement \emph{GreyConE} with state-of-the-art software testing tools: \ac{AFL} (v2.52b)\cite{afl}  and \ac{S2E}\cite{s2e}. For robust coverage measurements, we cross-validated our results with coverage measurement tools: \emph{lcov-1.13}~\cite{Lcov}, and \emph{gcov-7.5.0}~\cite{Gcov}. Experiments were on a 3.20 GHz 16 GB RAM $i5$ linux machine.

\begin{table}[t]
    \centering
    \setlength\tabcolsep{1.75pt}
    \caption{\emph{\solution{} execution phases}}
    \resizebox{\columnwidth}{!}{%
    \begin{tabular}{lr|rrc|rrr|rrr}
        \toprule
        \toprule
        \multirow{2}{*}{\textbf{Benchmarks}} &
        \multirow{2}{*}{\textbf{LOC}} &
        \multicolumn{3}{c|}{\textbf{\# Test-cases}} & \multicolumn{3}{c|}{\textbf{Branch cov. (\%)}} & \multicolumn{3}{c}{\textbf{Time(in s)}}\\
        \cmidrule(lr){3-5}\cmidrule(lr){6-8}\cmidrule(lr){9-11}
        & & {$fuzz_1$} & {$conc_1$} & {$fuzz_2$} & {$fuzz_1$} & {$conc_1$} & {$fuzz_2$} & {$fuzz_1$} & {$conc_1$} & {$fuzz_2$}\\ \midrule
        {ADPCM} & 270 & 5 & 6 & - & 93.3 & 100 & - & 9 & 39 & -  \\
        AES & 429 & 3 & 4 & 4 & 79.2 & 83.3 & 91.7 & 76 & 29 & 51 \\ 
        {FFT\_fixed} & 334 & 3 & 3 & - & 81.2 & 96.9 & - & 29 & 137 & - \\
        {IDCT} & 450 & 62 & 137 & - & 64.8 & 100 & - & 7 & 222 & - \\
        {MD5C} & 467 & 5 & 3 & 2 & 87.5 & 90.6 & 100 & 9 & 34 & 7 \\
        Filter\_FIR & 176 & 3 & 2 & 7 & 76.8 & 87.5 & 93.8 & 7 & 23 & 37 \\  
        {Interpolation} & 231 & 44 & - & - & 100 & - & - & 3 & - & - \\ 
        Decimation & 422 & 3 & 2 & - & 96.8 & 100 & - & 12 & 22 & - \\
        {Kasumi} & 415 & 31 & 23 & - & 93.3 & 100 & - & 10 & 46 & - \\
        UART & 160 & 2 & 15 & 4 & 81.2 & 84.1 & 88.5 & 334 & 136 & 248 \\
        {Quick\_sort} & 204 & 7 & - & - & 100 & - & - & 14 & - & - \\
        \bottomrule
        \bottomrule
    \end{tabular}
    }
    \label{tab:trojan_detection}
\end{table}

\subsection{Benchmark characteristics}

We evaluate \solution{} on a wide spectrum of available SystemC benchmarks: SCBench\cite{scbench} and S2CBench\cite{s2cbench}, selected from a wide variety of application domains covering many open-source hardware designs. The benchmarks considered have diverse characterization: ADPCM, FFT, IDCT (all image processing cores), AES, MD5C (cryptographic cores), Quick\_sort (data manipulation), Decimation (filters) and UART (communication protocols). 
\subsection{Design of experiments}

We evaluate the efficacy of \emph{GreyConE} in two aspects: (a) coverage improvement and (b) run-time speedup and compare our results with state-of-the-art testing techniques: fuzz-testing based approach (AFL) and concolic execution (S2E).

\textbf{Baseline 1 (Fuzz testing)}: We run \ac{AFL} on the benchmarks with default setting, and randomly generated initial seed inputs.

\textbf{Baseline 2 (Concolic execution)}: Similarly, we run \ac{S2E} on the benchmarks with default configurations and randomly generated concrete seed inputs.

\textbf{GreyConE}: We run \emph{GreyConE} on the SystemC benchmarks starting with randomly generated input test-cases. We set the user defined parameters $t_{threshold_f}=5$s and $t_{threshold_c}=10$s  for \ac{AFL}  and \ac{S2E} respectively, excluding the execution time needed to generate the first seed for each test engine.  



For our experiments, we set $time_{cutoff}$ to $2$ hours and compare baseline methods and \solution{}. Next, we discuss the performance of \solution{} in terms of branch-coverage improvement and run-time speedup.

\section{Results and evaluation}


 \emph{\solution{}} invokes fuzz-engine and concolic engine interchangeably. We have annotated each phase of run incrementally where $fuzz_{k}$ denotes $k^{th}$ execution phase of fuzz-engine. To show the effectiveness of \solution{}, we present branch coverage achieved in each phase along with a number of test-cases generated in \autoref{tab:trojan_detection}. 
 In~\autoref{tab:trojanDetectionGreyConE}, we report the number of test-cases generated; achievable branch coverage within $time_{cutoff}$ and the earliest time taken to reach that coverage. We compare our results with \ac{AFL}~\cite{afl} and \ac{S2E}~\cite{s2e} which are open-sourced implementations. Due to unavailability of implementations of \texttt{SESC}~\cite{2016_concolic} and \texttt{CTSC}~\cite{2018_concolic_sysC}, we used \ac{S2E}~\cite{s2e} to demonstrate the performance of concolic execution on SystemC designs by adopting necessary changes.

\begin{table}[!tb]
\centering
\caption{Comparing \emph{GreyConE}(GCE) with baselines
}
\setlength\tabcolsep{3pt}
\resizebox{\columnwidth}{!}{
\begin{tabular}{lr|rrr|rrr|rrrr}
\toprule
\toprule
\multicolumn{1}{c}{{\bf{Benchmarks}}} & \multicolumn{1}{c}{{\bf{LOC}}} & \multicolumn{3}{|c|}{\bf{\# Test-cases}} &
\multicolumn{3}{c|}{\bf{Branch cov.(\%)}} &  \multicolumn{3}{c}{\bf{Time taken (s)}} \\ 
\cmidrule(lr){3-5} \cmidrule(lr){6-8}\cmidrule(lr){9-11}
& & \begin{tabular}[c]{@{}l@{}}\bf{AFL}\end{tabular} &  \begin{tabular}[c]{@{}l@{}}\bf{S2E}\end{tabular} & \begin{tabular}[c]{@{}l@{}}\ \bf{GCE}\end{tabular} & 
\begin{tabular}[c]{@{}l@{}}\bf{AFL}\end{tabular} &  \begin{tabular}[c]{@{}l@{}}\bf{S2E}\end{tabular} & \begin{tabular}[c]{@{}l@{}}\ \bf{GCE}\end{tabular} & \begin{tabular}[c]{@{}l@{}}\bf{AFL}\end{tabular} &  \begin{tabular}[c]{@{}l@{}}\bf{S2E}\end{tabular} & \begin{tabular}[c]{@{}l@{}}\ \bf{GCE}\end{tabular} \\
\cline{1-11}

        {ADPCM} & 270 & 30 & 21 & {\bf 11}  & 96.9 & 94.4 & {\bf 100} & 124 & 374 & {\bf 48} \\
        AES & 429 & 23 & 8 & 11 & 88.7 & 82.4 & {\bf 91.7} & 1745 & 314 & {\bf 156}  \\  
        FFT\_fixed & 334 & 7 & 46 & {\bf 6} & 96.9 & 96.9 & 96.9 & 1215 & 2051 & {\bf 166} \\
        IDCT & 450 & 110 & 146 & 199 & 74.1 & 84.2 & {\bf 100} & 29 & 422 & 229 \\
        MD5C & 467 & 11 & 45 & {\bf 10} & 90.6 & 70 & {\bf 100} & 1329 & 1214 & {\bf 50} \\
        Filter\_FIR & 176 & 11 & 36 & 12 & 93.8 & 93.8 & 93.8 & 551 & 89 & {\bf 67} \\
        {Interpolation} & 231 & 44 & 5 & 44 & 100 & 100 & 100 & {\bf 3} & 37 & {\bf 3} \\
        Decimation & 422 & 4 & 5 & 5 & 100  & 100 & 100 & 372 & 1065 & {\bf 34} \\
        {Kasumi} & 415 & 60 & 76 & {\bf 54} & 100 & 100 & 100 & 58 & 233 & {\bf 56} \\
        UART & 160 & 4 & 54 & 21 & 81.2 & 85.7 & {\bf 88.5} & 334 & 730 & 718 \\
        {Quick\_sort} & 204 & {\bf 7} & 13 & {\bf 7} & 100 & 100 & 100 & {\bf 14}  & 248 & {\bf 14} \\
\bottomrule
\bottomrule
\end{tabular}
\label{tab:trojanDetectionGreyConE}
}
\end{table}

\begin{figure}
    \centering
\includegraphics[width=0.53\columnwidth]{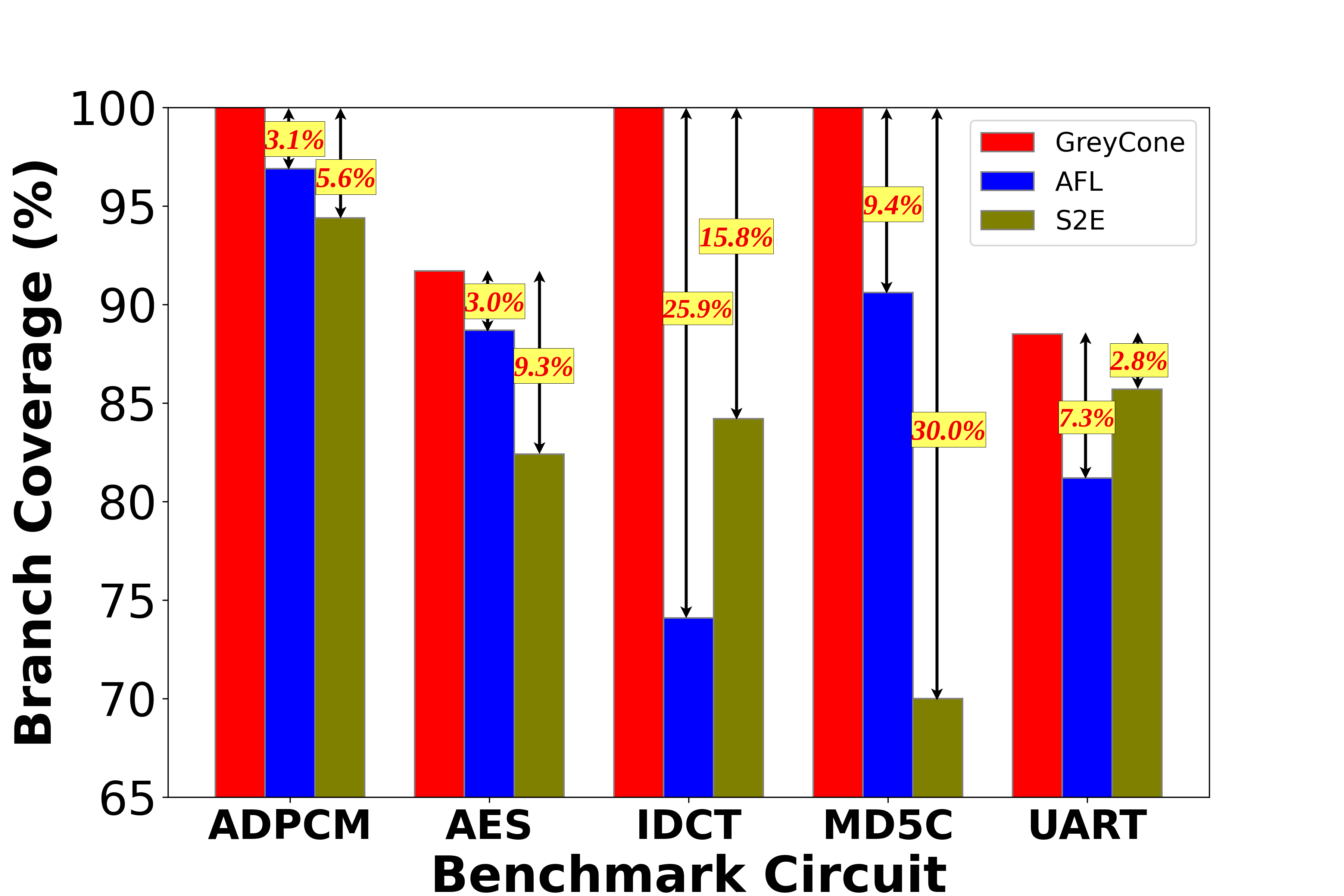}
\hspace{-20pt}
 \includegraphics[width=0.48\columnwidth]{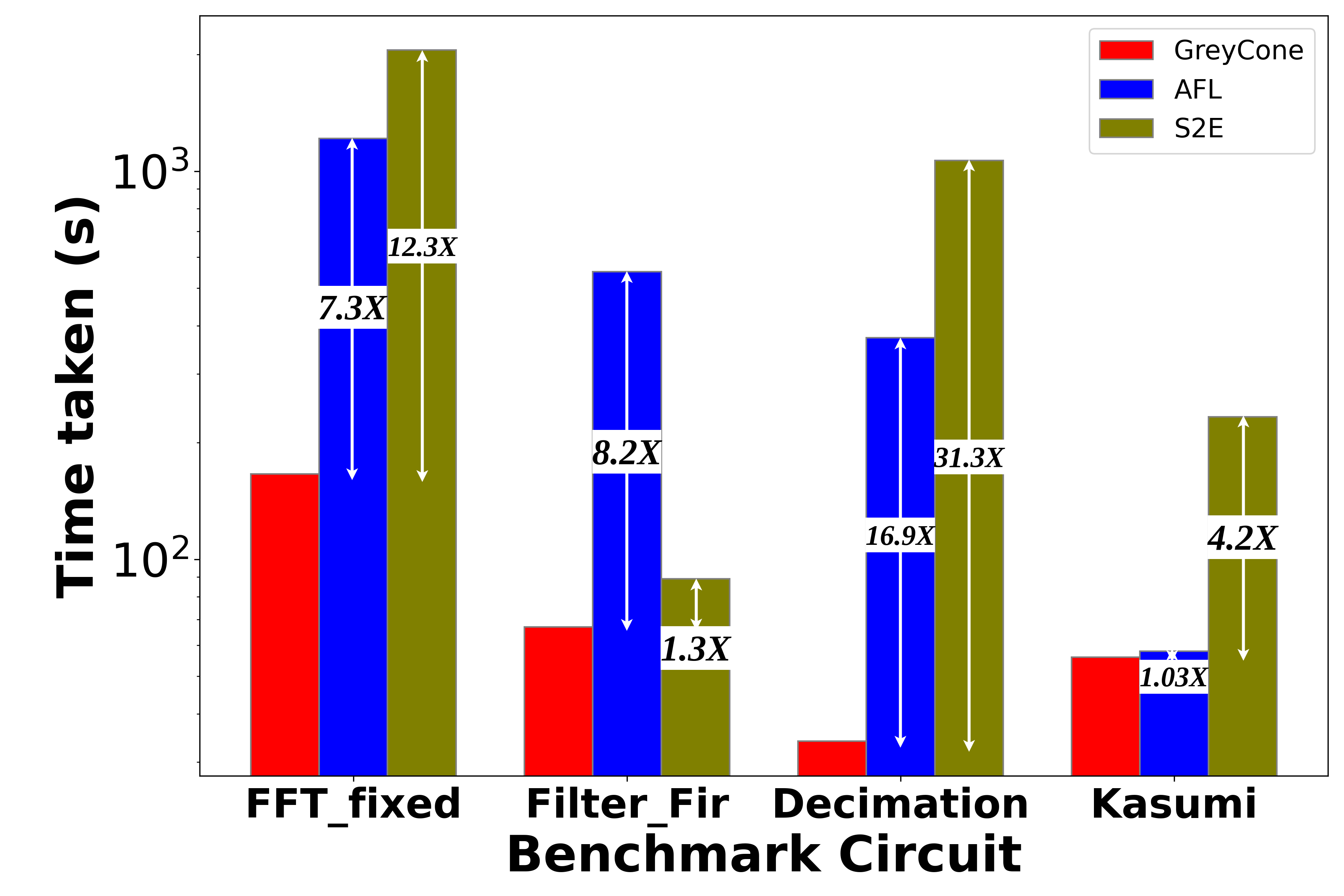}
 \caption{Coverage improvement (left) and timing speed-up (right) by \solution{}}
 \label{fig:solutionCoverageAndSpeedup}
\end{figure}


\input{images/fig_branchCov}



\noindent{\textbf{1) Coverage achieved by GreyConE:}} 
Higher branch coverage indicates greater probability of test-cases detecting bugs hidden in deeper program segments. As depicted in ~\autoref{tab:trojan_detection}, for certain designs, \solution{} does not require conocolic engine at all. This indicates that such benchmarks lack complex ``branching" conditions where fuzzer can get stuck. For every design, \solution{}  achieves maximum possible coverage post one call to concolic engine. Here, we claim \solution{} achieved ``maximum" possible coverage as we independently validated that ``uncovered" branches are unreachable codes (\textit{AES}, \textit{FFT\_fixed}, \textit{Filter\_FIR} and \textit{UART}). 
We compare the branch coverage obtained by \emph{GreyConE} with other techniques in \autoref{tab:trojanDetectionGreyConE}. We observe that \emph{GreyConE} outperforms baseline techniques \emph{AFL} and \emph{S2E} in terms of coverage achieved (\autoref{fig:solutionCoverageAndSpeedup}). As illustrated, the test-cases produced by \emph{GreyConE} significantly improves the branch coverage by 3\%-25.9\% compared to \ac{AFL} and 2.8\%-30\% compared to \ac{S2E} within the two hours' time limit. Fig.~\ref{fig:coverageOfGreyConE} provides a detailed coverage analysis over the entire time period of two hours.

\noindent{\textbf{2) Analyzing run-time speedup:}}
We measure the time taken to obtain best achievable coverage by baseline techniques and compare run-time speedup by \emph{\solution{}} to achieve the same. From \autoref{fig:coverageOfGreyConE}, the time taken to achieve a certain branch coverage is lower bounded by \emph{GreyConE} compared to \ac{AFL} and \ac{S2E}. In \autoref{fig:solutionCoverageAndSpeedup}, we show the run-time speedup of \emph{GreyConE} over \ac{AFL} and \ac{S2E} for the designs where every technique has achieved the same branch coverage within the time-limit. We observe that \emph{GreyConE} is significantly faster to reach a certain branch coverage. The speed-up achieved by \emph{GreyConE} are in line with our design approach: (1) \emph{GreyConE} quickly identifies the region where fuzz engine gets stuck and invoke concolic engine to solve the complex conditions; (2) \emph{GreyConE} avoids expensive path exploration by concolic execution by using fuzzer generated seeds leading to faster exploration and test-case generation.


\noindent{\textbf{3) Analyzing test-case quality:}} 
We report the number of test-cases preserved by each technique until it reaches user-defined timeout (or, till the maximum coverage is achieved) in \autoref{tab:trojanDetectionGreyConE}. A closer analysis reveal that the number of test-cases generated by \emph{GreyConE} is same as by \ac{AFL} where \ac{AFL} alone sufficed in reaching the target coverage without getting stuck for $t_{threshold_f}$ time. But, for cases where \ac{AFL} crossed the $t_{threshold_f}$, \emph{GreyConE} invokes concolic engine for generating quality test-cases quickly. Similarly when \ac{S2E} gets stuck for $t_{threshold_c}$, \emph{GreyConE} invokes the fuzz engine. Finally, \emph{GreyConE} needs fewer test-cases than both \ac{AFL} and \ac{S2E} indicating good quality test-case generation.

\label{label:exResults}

\section{Conclusion}
\label{sec:conclusion}
We proposed \solution{}: an end-to-end test-generation framework penetrating into deeper program segments of SystemC designs. Our results show scalable generation of test cases with better branch coverage and accelerated design space exploration compared to state-of-the-art techniques. \solution{} has alleviated the drawbacks of fuzzing and concolic execution by interleaving them. Future works include enhancing \emph{GreyConE} to low-level netlist (RTL/gate-level) and uncover hardware specific bugs in design.

\ifCLASSOPTIONcaptionsoff
  \newpage
\fi



%
\bibliographystyle{IEEEtran}
\bibliography{sample-base.bib}

\begin{thebibliography}{10}
\providecommand{\url}[1]{#1}
\csname url@samestyle\endcsname
\providecommand{\newblock}{\relax}
\providecommand{\bibinfo}[2]{#2}
\providecommand{\BIBentrySTDinterwordspacing}{\spaceskip=0pt\relax}
\providecommand{\BIBentryALTinterwordstretchfactor}{4}
\providecommand{\BIBentryALTinterwordspacing}{\spaceskip=\fontdimen2\font plus
\BIBentryALTinterwordstretchfactor\fontdimen3\font minus
  \fontdimen4\font\relax}
\providecommand{\BIBforeignlanguage}[2]{{%
\expandafter\ifx\csname l@#1\endcsname\relax
\typeout{** WARNING: IEEEtran.bst: No hyphenation pattern has been}%
\typeout{** loaded for the language `#1'. Using the pattern for}%
\typeout{** the default language instead.}%
\else
\language=\csname l@#1\endcsname
\fi
#2}}
\providecommand{\BIBdecl}{\relax}
\BIBdecl

\bibitem{FormVerPetriNet}
D.~Karlsson \emph{et~al.}, ``Formal verification of systemc designs using a
  petri-net based representation,'' in \emph{DATE}, 2006, pp. 1228–--1233.

\bibitem{ModCheckSysCAuto}
P.~Herber \emph{et~al.}, ``Model checking systemc designs using timed
  automata,'' in \emph{CODES+ISSS}, 2008, pp. 131–--136.

\bibitem{symba}
A.~Vafaei \emph{et~al.}, ``Symba: Symbolic execution at c-level for hardware
  trojan activation,'' in \emph{ITC}, 2021, pp. 223--232.

\bibitem{Rolf1}
V.~Herdt \emph{et~al.}, ``Verifying systemc using stateful symbolic
  simulation,'' in \emph{DAC}, 2015, pp. 1--6.

\bibitem{DART}
P.~Godefroid \emph{et~al.}, ``Dart: Directed automated random testing,'' in
  \emph{PLDI}, 2005, pp. 213–--223.

\bibitem{sage}
P.~Godefroid, ``{SAGE}: Whitebox fuzzing for security testing,'' \emph{Queue},
  vol.~10, 2012.

\bibitem{afl}
M.~Zalewski, ``American fuzzy lop,'' \url{http://lcamtuf.coredump.cx/afl}.

\bibitem{snpfuzzing}
T.~Trippel \emph{et~al.}, ``Fuzzing hardware like software,'' \emph{IEEE S\&P},
  2021.

\bibitem{2018_concolic_sysC}
B.~Lin \emph{et~al.}, ``Concolic testing of systemc designs,'' in \emph{ISQED},
  2018, pp. 1--7.

\bibitem{s2e}
V.~Chipounov \emph{et~al.}, ``The s2e platform: Design, implementation, and
  applications,'' \emph{ACM TCS}, vol.~30, 2012.

\bibitem{2016_concolic}
B.~Lin \emph{et~al.}, ``Generating high coverage tests for {SystemC} designs
  using symbolic execution,'' in \emph{ASPDAC}, 2016, pp. 166--171.

\bibitem{chowdhury2019verifuzz}
A.~B. Chowdhury \emph{et~al.}, ``Verifuzz: Program aware fuzzing,'' in
  \emph{TACAS}, 2019, pp. 244--249.

\bibitem{LLVM}
``{LLVM},'' \url{https://llvm.org/}.

\bibitem{qemu}
F.~Bellard, ``{QEMU}, a fast and portable dynamic translator.'' in \emph{USENIX
  FREENIX}, 2005, p.~46.

\bibitem{klee}
C.~Cadar \emph{et~al.}, ``{KLEE}: unassisted and automatic generation of
  high-coverage tests for complex systems programs,'' in \emph{OSDI}, 2008.

\bibitem{Lcov}
``{LCOV},'' \url{http://ltp.sourceforge.net/coverage/lcov.php}.

\bibitem{Gcov}
``{GCov},'' \url{https://gcc.gnu.org/onlinedocs/gcc/Gcov.html}.

\bibitem{scbench}
B.~Lin \emph{et~al.}, ``Scbench: A benchmark design suite for systemc
  verification and validation,'' 2018.

\bibitem{s2cbench}
B.~C. Schafer \emph{et~al.}, ``S2cbench: Synthesizable systemc benchmark suite
  for high-level synthesis,'' \emph{IEEE ESL}, vol.~6, 2014.

\end{thebibliography}

%








\end{document}